\documentclass[aps,prl,reprint,groupedaddress,showpacs,floatfix]{revtex4-1}

\usepackage{graphicx}
\usepackage{dcolumn}
\usepackage{bm}

\usepackage{graphicx}
\usepackage{amssymb,amsfonts,amsmath}
\usepackage{framed}

\begin{document}


\title{An action principle of classical irreversible thermodynamics -\\ Irreversible thermodynamic cycles and embodied bits of information}


\author{Rudolf Hanel}
\email[]{rudolf.hanel@meduniwien.ac.at}

%
\affiliation{Section for Science of Complex Systems, Medical University of Vienna, Spitalgasse 23, 1090 Vienna, Austria}


\date{\today}

\begin{abstract}
Despite its simplicity, it seems to my best of knowledge that the possibly simplest approach towards deriving 
equations governing irreversible thermodynamics from gas-kinetic considerations within the framework of classical mechanics 
has never been pursued. In this paper we address this omission and derive the equations 
describing the irreversible thermodynamics of a gas in a piston and associated thermodynamic cycles performed in finite time.
What we find is a thermodynamic action principle: The irreversible work we require for performing a thermodynamic cycle in finite time
times the time we require to run through the cycle, a isothermal compression/decompression cycle for instance,
will always be larger or equal to a lower bound given by a system specific constant with the dimension of an action. 
This process specific action constants can take values of the order of Plank's constant for microscopic processes, such as displacing a 
Hydrogen atom by one atom diameter. For macroscopic processes (e.g. a bicycle pump) also the action constant takes macroscopically
observable values. We discuss the finite time Carnot cycle, and estimate the lower bound for irreversible 
work required when rewriting physically embodied bits of information and compare the resulting values with Landauer's bond, 
the irreversible amount of energy that needs to be dissipated for re-writing a bit. 
\end{abstract}


\maketitle

\section{Introduction}

Various approaches have been pursued in order to understand and develop mathematical frameworks for 
describing phenomena of non-equilibrium thermodynamics, \cite{netd}. Phenomena, such as for instance coupled 
transport processes \cite{Onsager}, or the thermodynamics of finite-time Carnot-cycles \cite{CurzonAhlborn}. 
{\em Classical irreversible thermodynamics} (CIT), for instance, is using the so called local equilibrium 
assumption \cite{Onsager,Prigogine,HafskoldKjelstrup}, which is also underlying {\em superstatistics} \cite{BeckCohen,HTMGM1}, 
assuming that processes consisting of an extended system of interacting entities, such as the molecules in a fluid, 
can be described as being locally in equilibrium. 
The local equilibrium hypothesis however may not always hold. particularly if a systemic energy potential, 
e.g. two heat bathes, drives a continuous energy current through the system.

To describe complex non-equilibrium situations approaches other than CIT get considered. 
This includes direct simulation of systems, which typically requires other
mathematical and computational means than local equilibrium theories, e.g. \cite{Chandrasekhar,CrossHohenberg}. 
Here we consider {\rm thermodynamic irreversibility} as a consequence of performing thermodynamic cycles in 
finite time. It is surprising to notice that despite the success of gas-kinetic considerations
in describing the quasi-stationary (local equilibrium) thermodynamic laws of the ideal-gas 
in the early days of an emergent statistical theory of physics, \cite{Clausius,Maxwell}, these considerations 
(to my best knowledge) have never been extended to derive a corresponding theory for systems driven irreversibly in finite time.
In the following we will remedy this omission. 
To do so, we consider a model where a cylinder is filled with an ideal gas, i.e. a collection of point particles
that elastically collide with a piston moving at finite velocities according to the laws of classical mechanics. 

\begin{figure}[th]
	\begin{center}		
		\includegraphics[width=0.9\columnwidth]{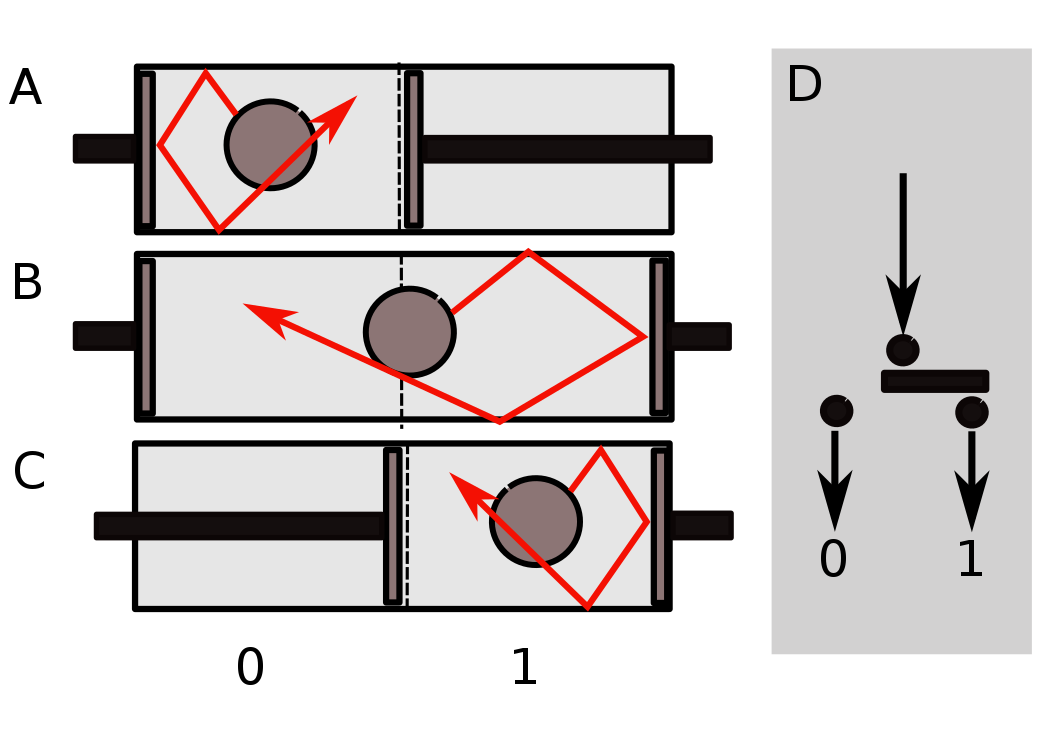}
	\end{center}		
	\caption{Thermodynamic model of resetting a bit. (A) A single particle is confined in volume $0$ by two pistons.
	(B) One piston is retracted and the particle can now occupy volume $0$ and $1$. (C) The second piston confines
	the particle in volume $1$. Reseting the bit, i.e. pushing the particle from volume $0$ to volume $1$, 
	therefore is equivalent to a thermodynamic compression-cycle, that, in case the process is coupled to a heat-bath, 
	corresponds to an isothermal expansion followed by an isothermal compression of the system.  
	(D) A symbolic switch for setting a bit of information.  
	}
	\label{fig0}
\end{figure}

What we find is an action principle: $\tau_{\rm cycle}\Delta W_{\rm irrev}\geq Nh_{\rm cycle}$. 
The time $\tau_{\rm cycle}$ it takes to run through a thermodynamic cycle times the amount of irreversible work 
per molecule $\Delta W_{\rm irrev}/N$ that is required to run the thermodynamic cycle is always larger than a 
system specific constant, the action $h_{\rm cycle}$. 
This irreversible action only depends on geometric parameters, such as extremal piston positions in the cycle, and the temperatures of 
heat-baths, defining the conditions under which the process performs. As a consequence of this action principle it becomes    
clear that in the limit $\lim_{\tau_{\rm cycle}\to\infty} \Delta W_{\rm irrev} = Nh_{\rm cycle}/\tau_{\rm cycle} \to 0$. In other words, the equations of
(quasi-stationary) reversible processes are recovered in the static limit $\tau_{\rm cycle}\to\infty$.
We will use the relations obtained for finite time cycles to estimate the efficiency of the irreversible Carnot cycle.
Yet, the even simpler compression-expansion cycles are important in themselves, accompanying every process of one body 
pushing another one. Moreover we can use those simple cycles to estimate the power consumption necessary for rewriting 
embodied bits of information.
Such bits could be any physical device with two distinct states, compare Fig. \ref{fig0}D). 
In a macroscopic device, such as a light switch, switching between binary states may simply mean moving a piece of copper rod. 
In a microscopic device one might think of relocating a single molecule or atom by a distance of the magnitude of the molecule diameter.
However, pushing a molecule from a location $0$ into an adjacent location $1$ (compare Fig. \ref{fig0}A-C) can be regarded
as a thermodynamic cycle where an expansion phase is followed by a compression phase, and
we can use ``Szilard's one particle gas'' model, \cite{Szilard}, to represent one bit of physically embodied information. 
 
Two fundamental properties of classical computation devices are (i) how much energy do we have to dissipate in order to fix the value 
of a bit reliably after resetting a bit 
and (ii) how much irreversible work do we need to invest into resetting the bit in finite time.
The first quantity gets estimated by the lower energy bound, $\Delta W_{\rm Landauer}=kT\log 2$, known as
Landauer's bound \cite{Landauer}, which also forms a lower bound on the hight of the energy potential
separating the two states of the embodied bit. The second quantity is a consequence of the properties of 
an ideal gas being used in irreversibly driven thermodynamic cycles.
In the following we will derive the corresponding equations.

\section{A gas kinetic approach to irreversible thermodynamic cycles}
 
\begin{figure}[t]
	\begin{center}		
		\includegraphics[width=0.85\columnwidth]{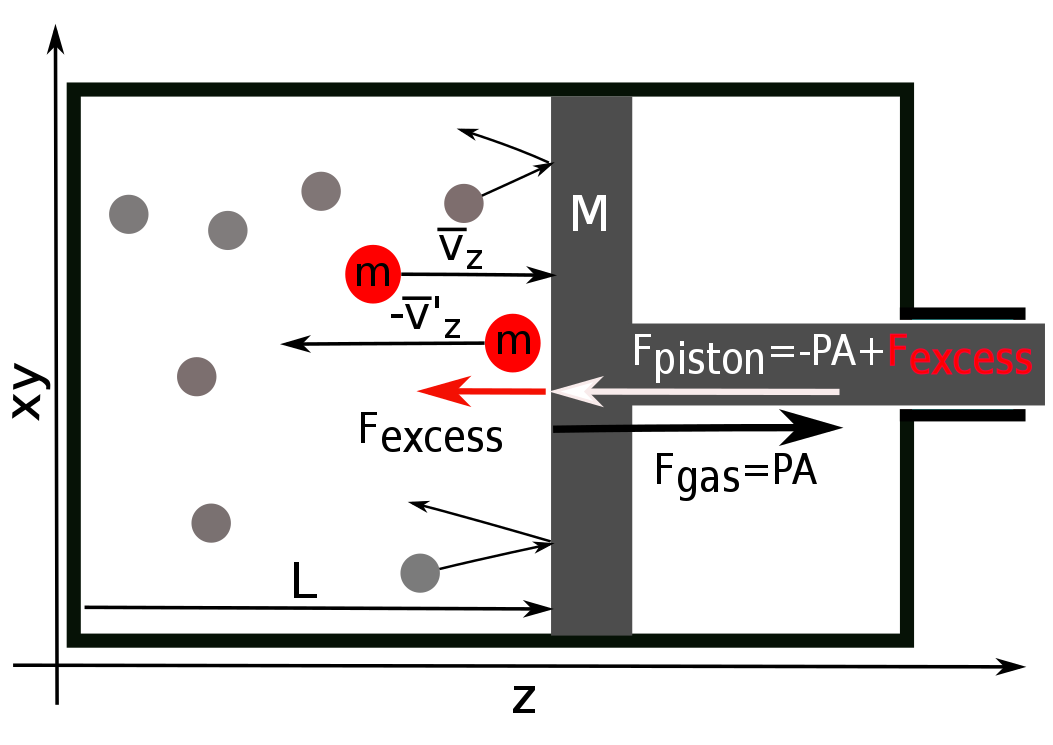}
	\end{center}		
	\caption{Irreversible processes are also {\em non-equilibrium} processes. Irreversible here means that the force used to push the piston $F_{\rm piston}$
	does not merely compensate the force $F_{\rm gas}$ the gas exerts on the piston. The excess forces $F_{\rm excess}=F_{\rm gas}+F-{\rm piston}$ 
	drives the system in finite time but also breaks homogeneity and isotropy of the gas-particle distribution in the cylinder, however slightly.
	We will assume that the excess force is small enough for the gas to remain homogeneously distributed in the cylinder, except maybe 
	for molecules in close proximity of the piston surface after they collided with the piston. In this region the velocity distribution of 	
	gas-molecules gets skewed between particles that approach, and those which rebound from the piston!
	}
	\label{fig1}
\end{figure}

Let us consider a system of $N$ particles (gas-molecules) with $f=3$ degrees of freedom!
The molecules are contained in a cylinder with a cross-section $A$ ($xy$-plane of our coordinate system, compare Fig. \ref{fig1}), 
and a piston that allows us to vary the hight $L$ of the cylinder ($z$-direction of the coordinate system). 
The cylinder volume is given by $V=AL$, where $L$ is the length of the gas filled cylinder volume controlled by the piston. 
The piston velocity $u=\dot L$ is negative for compression and positive for expansion.
  
The average velocity of a particle is given by $\bar v=\langle |v|^2\rangle^{1/2}>0$. Each particle has a
velocity $v_i$ ($i=1,\cdots,N$). Homogeneity and isotropy imply that the gas on average behaves identical 
everywhere and in all directions!
In this case we find that
\begin{equation}
\bar v^2=\langle v_x^2\rangle + \langle v_y^2\rangle + \langle v_z^2\rangle = 3 \langle v_z^2\rangle
\end{equation}
In particular homogeneity implies that $\bar v$ does not depend on the position $x$ in the cylinder. 
Isotropy on the other hand implies that $\langle v_x^2\rangle = \langle v_y^2\rangle = \langle v_z^2\rangle$.
Homogeneity and Isotropy of gases in equilibrium are a consequence of the equipartition 
principle of energy! 
\\

Let us ask now the following question.
How much work $W$ do we need to compress the system from volume $V_1$ to $V_2$, disregarding friction of the piston with the cylinder?
I.e. we are not interested in irreversible work due to friction but merely due to finite time processing.
For compressing the gas in the cylinder from volume $V_1$ to $V_2<V_1$ we have to push the piston against the force
exerted from molecules rebounding from the piston and then we need a little extra force $F_{\rm excess}$ 
not only to hold but actually move the piston. In the end $F_{\rm excess}$ is responsible for the irreversible work that 
we have to invest into moving the piston. We further assume that the velocity of the piston $u$ is small with respect to the particle velocities $v$,
so that the change in kinetic energy of molecules after colliding with the piston can dissipate over the degrees of freedom of the 
gas molecules in the cylinder; and for isothermal processes equilibrate with the heat bath. 

Moreover, $|u|\ll |v|$ also guarantees that during compression (or expansion) the distribution of gas particle 
in the cylinder and their velocities remains homogeneous (except maybe in close vicinity of the piston).

From this assumption it follows that $1/2$ of all particles will on average move towards the piston 
with average velocity $\bar v_z=\sqrt{\langle v_z^2\rangle}>0$. The average distance $\Delta x_z$  
between particles in $z$-direction (the direction normal to the piston's surface) that move toward the piston 
is given by
\begin{equation}
\Delta x_z=L/(N/2)=\frac{2L}{N}\,.
\label{delx}
\end{equation}
Moreover, the average time $\Delta t_z$ elapsing between subsequent collisions of particles with the piston surface 
becomes
\begin{equation}
\Delta t_z=\Delta x_z/(\bar v_z - u)=\frac{2L}{N(\bar v_z - u)}
\label{delt}
\end{equation}

In the next step we look at the collision processes between particles and piston, where
$m$ is the particle mass and $M$ is the mass of the piston.
$\bar v_z$ and $\bar v_z'$ are the average particle velocities,  
and $u_-$ and $u_+$ the piston velocity before and after the collision. 
We note that by convention we assume the velocities $\bar v_z>0$ and $\bar v_z'>0$. 
In our coordinate system a particle moves towards the piston with velocity $\bar v_z$ and 
rebounds from the piston with velocity $-\bar v_z'$.
Therefore momentum conservation implies
\begin{equation}
m \bar v_z +\quad M u_- = -m \bar v_z' +\quad M u_+
\label{momentumcons}
\end{equation}

\begin{figure}[t]
	\begin{center}		
			\includegraphics[width=0.85\columnwidth]{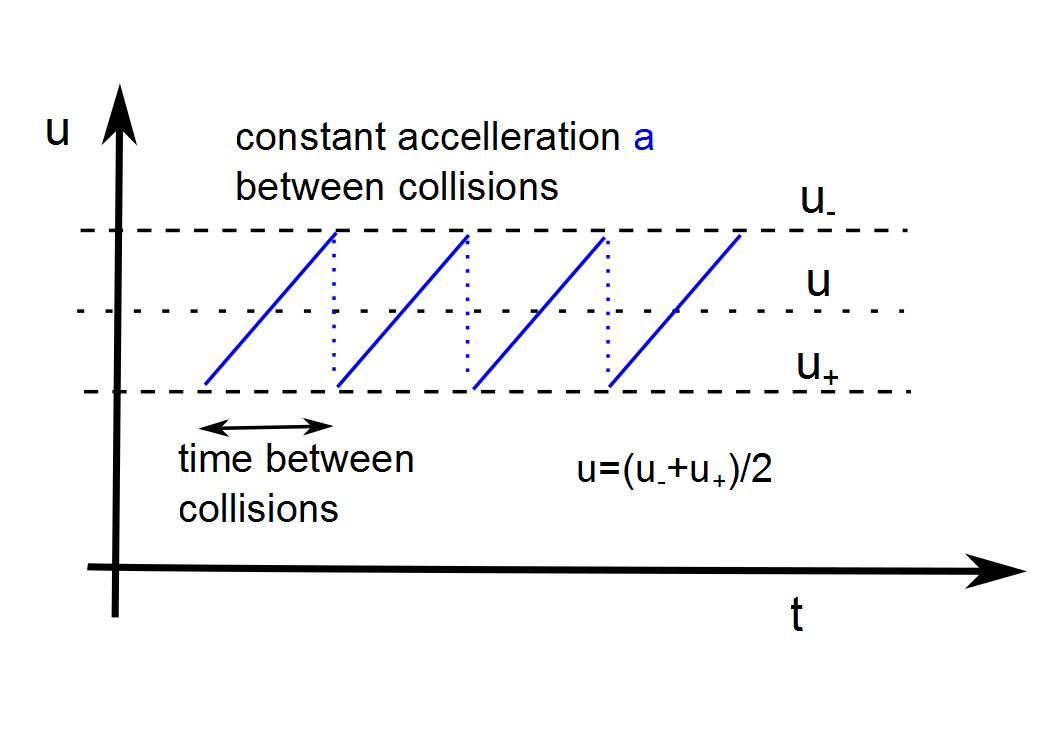}
	\end{center}		
	\caption{The time evolution of the piston velocity is shown for a particle colliding 
	with the piston every $\Delta t_z$ time units and the piston maintaining an average velocity $u$ always 
	re accelerating from $u^+$ after a collision event, to $u_-$ directly before the next collision.
	}
	\label{fig2}
\end{figure}

From energy and momentum conservation we get 
\begin{equation}
u_+= \frac{M-m}{M+m} u_- + \frac{2m}{M+m}\bar v_z
\label{collisionA}
\end{equation}
and 
\begin{equation}
\bar v'_z= \frac{M}{m}(u_+ - u_-) - \bar v_z
\label{collisionB}
\end{equation}

The piston velocity $u$ can only be an average velocity. 
In order to maintain the average piston velocity $u$ we require
\begin{equation}
u= \frac{u_- + u_+}{2}\,.
\label{uavg}
\end{equation}
Also, between collisions the piston re-accelerates from $u_+$ back to $u_-$ with a (constant) acceleration $a$. 
From $u_-= u_+ + a \Delta t_z$ it follows that
\begin{equation}
u_\pm=u \mp \frac{1}{2}a\Delta t_z
\label{upmis}
\end{equation}
and
\begin{equation}
a= -\left(\bar v_z - u\right)^2\frac{mN}{ML}\,.
\label{accel}
\end{equation}
Note that the sign of the acceleration is always negative, irrespective whether $u$ is positive and the volume in the piston increases
or $u$ is negative and the volume decreases. Using that the force $F_z$ required to accelerate $M$ with $a$ is $F_z=Ma$, the gas pressure against the moving 
piston is $P_z=-F_z/A>0$ while the gas pressure against the static walls of the cylinder 
is given by $P=-F/A$ with $F=a_0M$ and $a_0=\bar v_z^2\frac{mN}{ML}$. 
The volume in the cylinder is $V=AL$ and one gets 
\begin{equation}
P_zV= mN \left(\bar v_z - u\right)^2
\label{equstateA}
\end{equation}

We recall the assumption we made that $|u|$ is small enough for the gas to remain homogeneous during compression.
This means that, as usual, we still can identify the temperature $kT/2$ with the {\em energy per degree of freedom} $f$. 
In particular we have $\bar v_x=\bar v_y=\bar v_z$ (except maybe very close to the piston surface) and therefore
\begin{equation}
kT = m (\bar v_z)^2=\frac{2}{f}\frac{U}{N}\,,
\label{tempdef}
\end{equation}
where $U$ is the {\em internal energy} of the system.
For the mono-atomic gas we have $f=3$ and $U=E_{\rm kin}=mv^2/2=3mv_z^2/2$.
For the static case $u=0$ we have $P=P_z$ and Eq. (\ref{equstateA}) turns into the equation of states for the ideal gas
\begin{equation}
PV= NkT
\label{equstateB}
\end{equation}


\subsection{(A) Isothermal irreversible work}

How much work do we need for compressing the gas from $V_1$ to $V_2<V_1$. 
This work is
\begin{equation}
dW=F_zdL=MadL=-\left(1-\sqrt{\frac{m}{kT}}u\right)^2NkT\frac{dL}{L}\,,
\label{isowork}
\end{equation}
and therefore
\begin{equation}
\Delta W|_1^2=-\left(1-\sqrt{\frac{m}{kT}}u\right)^2NkT\log\frac{V_2}{V_1}\,,
\label{isowork2}
\end{equation}
If we perform the cycle of a isothermal compression from $V_1$ to $V_2$ and then a expansion from $V_2$
back to $V_1$, then the irreversible work $\Delta W_{\rm isothermal}=\Delta W|_1^2+\Delta W|_2^1$ performed 
during the cycle is given by
\begin{equation}
\Delta W_{\rm isothermal}=4N|u|\sqrt{mkT}\log\frac{V_1}{V_2}>0\,.
\label{isoworkcyc}
\end{equation}
The cycle gets performed in the cycle-period
\begin{equation}
\tau_{\rm cycle}=\frac{2}{|u|}(L_1-L_2)>0\,.
\label{tcycle}
\end{equation}
The {\em action} per particle per cycle $h_{\rm isothermal}\equiv\tau_{\rm cycle}\Delta W_{\rm isothermal}/N$ 
is a constant depending only on the particle mass $m$ the temperature $T$ and the lengths $L_1$ and $L_2$:
\begin{equation}
h_{\rm isothermal}=8\sqrt{mkT}(L_1-L_2)(\log L_1-\log L_2)>0\,.
\label{isoactcyc}
\end{equation}

The irreversible power consumption of an isothermal cycle becomes
\begin{equation}
{\cal P}_{\rm isothermal}=N\frac{h_{\rm isothermal}}{\tau_{\rm cycle}^2}>0\,,
\label{isopowcyc}
\end{equation}
which diverges like $\tau_{\rm cycle}^{-2}$ as $\tau_{\rm cycle}\to0$.
Note that the positive value of ${\cal P}_{\rm isothermal}$ means that dissipative work has been performed on the system
and that this energy is irreversibly lost in form of heat absorbed by the heat bath. 
 
\subsection{(B) Adiabatic irreversible work}

In contrast to the isothermal cycle the adiabatic cycle is isolated and
no heat is transfered to the heat bath but work performed on the system solely increases the
internal energy $U$ of the system. Work performed on the system $dU=dW$ increases the kinetic energy of the gas particles in the cylinder
and the internal energy now depends on the position of the piston $L$.
We therefore substitute $U\to U_+(L)$ during compression ($dL<0$) and $U\to U_-(L)$ during expansion ($dL>0$). 

Again, if $|v_z|\gg|u|$, then the gas (except maybe a thin layer close to the piston) will essentially 
remain homogeneous during adiabatic compression and we can still use the usual temperature definition in the bulk of the gas, i.e.
\begin{equation}
kT =\frac{2}{fN}U_{\pm}(L)\,,
\label{temp}
\end{equation}
where $f$ is the number of degrees of freedom of a molecule ($f=3$ for a mono-atomic gas) and 
$U_{\pm}(L)=fN\frac12m\bar v_z(L)^2$.
With $dW=MadL=-(\bar v_z-u)^2mNdL/L$ we get 
\begin{equation}
\frac{dT}{T\left(1-\sqrt{\frac{m}{kT}}u\right)^2}=-\frac2f\frac{dL}{L}\,.
\label{adaibat3}
\end{equation}
Using the variable transformation $x=\sqrt{kT}-\sqrt{m}u$ allows us to integrate the equation and obtain:
\begin{equation}
\log\left(\sqrt{kT}-u\sqrt{m}\right)-\frac{\sqrt{\frac{m}{kT}}u}{1-\sqrt{\frac{m}{kT}}u}+\frac1f\log L={\rm const.}\,.
\label{adiabatInt}
\end{equation}
For obtaining an explicit expression for $T(L_2)$, in the integral Eq. (\ref{adiabatInt}), 
we will expanding this expression to the first order in $u\sqrt{\frac{m}{kT}}$. This is justified 
since e.g. the proton-mass is of order $10^{-27}$ Kg and the Boltzmann constant $k$ is of order $10^{-23}$ J/K. 
From this it follows that $\sqrt{\frac{m}{kT}}$ is typically of the order $10^{-3}$ s/m for room temperature and 
we get 
\begin{equation}
\log\frac{T_2}{T_1}-4u\sqrt{\frac{m}{k}}\left(\frac1{\sqrt{T_2}}-\frac1{\sqrt{T_1}}\right)=-\frac2f\log\frac{L_2}{L_1}\,.
\label{intApprox}
\end{equation}
Using $a=4u\sqrt{m/k}$ and $b=\log T_1-4u\sqrt{m/kT_1}-(2/f)\log(L_2/L_1)$
we can rewrite Equation (\ref{intApprox}) into 
\begin{equation}
\log T_2 - \frac{a}{\sqrt{T_2}}=b\,,
\label{intApprox2}
\end{equation}
which can be solved using the Lambert W-function, $W(x\exp(x))=x$:
\begin{equation}
T_2 = \left(\frac{2}{a}W\left(\frac{a}{2}e^{-\frac{b}{2}}\right)\right)^{-2}\,.
\label{solutionApprox}
\end{equation}
Using that $W(x)\sim x$ for $x\sim 0$ we can derive the following approximate solution:  
\begin{equation}
T_2 \sim e^{b} = T_1\left(\frac{L_2}{L_1}\right)^{-\frac{2}{f}}e^{-4u\sqrt{\frac{m}{kT_1}}}\,.
\label{solutionApprox2}
\end{equation}
If we go through the irreversible cycle $1\to2\to3$ with $L_3=L_1$, 
where we compress adiabatically ($1\to 2$ with $u<0$) and decompress adiabatically 
($2\to 3$ with $u>0$), we can approximately compute $\Delta T=T_3-T_1$ up to an order ${\cal O}(u^2m/kT)$. 
We obtain for $L_2<L_1$, and $T=T_1$, that
\begin{equation}
\frac{\Delta T}{T} \sim -4u\sqrt{\frac{m}{kT}}\left(1-\left(\frac{L_2}{L_1}\right)^{\frac{1}{f}}\right)>0\,.
\label{DeltaTApprox}
\end{equation}
The temperature in the piston at the end of the adiabatic cycle has increased by $\Delta T$.
Since the temperature of the system has increased, repeating the cycle requires to first
cool down the system from $T+\Delta T$ to $T$ again, otherwise each adiabatic cycle will 
successively increase the temperature of the system, corresponding to increasing the internal energy $U$ by 
the amount of irreversible work required for running the cycle in finite time, which has dissipated over the degrees of freedom of the system.

Analogous to the isothermal case we can define an {\em action} per particle for
the adiabatic cycle $h_{\rm adiabatic}=\tau_{\rm cycle}\Delta W_{\rm adiabatic}/N$,
where $\Delta W_{\rm adiabatic}$ is the dissipative work required for performing an adiabatic 
compression cycle within the time $\tau_{\rm cycle}$. Within the accuracy of our approximations we find:
\begin{equation}
h_{\rm adiabatic}=4f\sqrt{mkT}(L_1-L_2)\left(1-\left(\frac{L_2}{L_1}\right)^{\frac{f}{2}}\right)>0
\label{adiabatic-action}
\end{equation}
Again the adiabatic action is a positive constant, that does not depend on the cycle period $\tau_{\rm cycle}$, i.e. 
we loose the irreversible energy $\Delta W_{\rm adiabatic}=Nh_{\rm adiabatic}\tau_{\rm cycle}$ to the system 
by increasing its internal energy. 
Moreover, the adiabatic action always is larger than the isothermal action, i.e.:
\begin{equation}
h_{\rm adiabatic}>h_{\rm isothermal}\,,
\label{action-quot}
\end{equation}
i.e. if we run through the adiabatic cycle as quickly as through the analogous isothermal one, then we
require more irreversible work in the adiabatic case than we do in the isothermal one. 

\section{Discussion}

These simple gas-kinetic considerations show that the second law of thermodynamics in finite time thermodynamics 
relates to a thermodynamic action principle: 
\begin{framed}
{\em For any thermodynamics cycle there exists a constant action  $h_{\rm cycle}>0$ that does not depend on the 
numbers of particles $N$ or the cycle period $\tau_{\rm cycle}$, such that the dissipative irreversible work $\Delta W_{\rm irrev}$ 
is bounded from below, i.e. $\Delta W_{\rm irrev}\geq N h_{\rm cycle}/\tau_{\rm cycle}$.}
\end{framed}
The action $h_{\rm cycle}$ only depends on geometric variables such as $L_1$ and $L_2$, the particle mass 
$m$, the degrees of freedoms $f$, and most importantly one or more reference temperatures $T_i$ (e.g. the heat bath temperature).

\subsection{The finite time Carnot cycle}

We can use this action principle to estimate the efficiency of a finite time Carnot cycle. Further we can estimate the
irreversible power-consumption involved in switching bits, i.e. the lower bounds for the irreversible power-consumption of computing.
Let us consider a finite time Carnot cycle $1\to 2\to 3\to 4\to 1$, where $1\to 2$ corresponds to the isothermal compression from $L_1\to L_2$
coupled to a heat bath with temperature $T_1$, $2\to 3$ is an adiabatic compression from $L_2\to L_3$, where the temperature in the system increases from
$T_1\to T_2$. Then, the cycle expands isothermally from $L_3\to L_4$ coupled to a heat bath of temperature $T_2$ and then we adiabatically expand from
$L_4\to L_1$ such that at the end of the expansion we reach $T_1$ again.

If we require that the adiabatic work performed in the cycle cancels each other, as it does in the static case $\tau_{\rm cycle}\to\infty$, 
i.e. $W_2^3+W_4^1=0$, with the average piston velocity $\bar u=|u|$,
we find the following condition
\begin{equation}
\log\left( \frac{L_3}{L_2} \right)=\log\left( \frac{L_4}{L_1} \right)-2f\bar u\sqrt{\frac{m}{k}}\left(\frac1{\sqrt{T_2}}-\frac1{\sqrt{T_1}}\right)\,.
\end{equation}
This condition allows us to continue to isothermally compress the system directly after the adiabatic expansion has finished 
at a temperature matching exactly the temperature $T_1$ of the heat bath. 
It further follows that the finite time Carnot efficiency is again given by $\eta=1-|W_1^2/W_3^4|$.
This implies that 
\begin{equation}
\eta=1-\gamma\frac{T_1}{T_2}
\label{carnoteta}
\end{equation}
where $\gamma$ is a factor arising from the Carnot cycle being performed in finite time.
Within the accuracy of the approximations we introduced for deriving the formulas for the adiabatic work it follows that:
\begin{equation}
\gamma=\frac{ \left( \frac{ 1+\bar u \sqrt{ \frac{m}{kT_1} } }{ 1-\bar u \sqrt{ \frac{m}{kT_2} }  }\right)^2 }
{1-2f\bar u \frac{\sqrt{ \frac{m}{k} }\left( \frac1{\sqrt{T_2}}-\frac1{\sqrt{T_1}} \right)}{\log L_2- \log L_1 }}\,.
\label{carnotgamma}
\end{equation}
Obviously, in the limit $\bar u\to 0$ we get $\gamma\to 1$ and we recover the reversible Carnot efficiency while for $\bar u >0$
the efficiency gets smaller.
Moreover, this result indicates that there exists a minimal cycle time $\hat \tau_{\rm cycle}$ corresponding to a maximal average piston speed
$\hat u$ (compare Eq. (\ref{tcycle})) such that the Carnot efficiency vanishes, i.e. $\eta(\hat u)=0$. 
This means that $\hat u$, or alternatively $\hat\tau$, corresponds to the {\em idle speed} of the finite time Carnot engine.
In this case the entire energy consumption of the process is spent on nothing else than driving the engine without exterior load.
In other words all the work performed is irreversible work.

\subsection{Rewriting bits}

By speeding up compression/decompression cycles the amount of energy required for irreversible work can become arbitrarily large
and therefore can become much larger than Landauer's bound. Since Landauer's bound and the bound we derived here rely on different 
physical reasons the total irreversible work required per bit and rewriting cycle needs to be larger than the sum of both contributions:
\begin{equation} 
\Delta W_{\rm irreversible}=\underbrace{kT\log2}_{\rm Landauer's\ bound}+\underbrace{\nu h_{\rm cycle}/\tau_{\rm cycle}}_{\rm finite\ time\ cycle}\,,
\end{equation} 
where $\nu$ is the number of particles per bit. For Szillard's one particle gas $\nu=1$.
The total minimal irreversible energy consumption for computing therefore can not be discussed independently from the frequency of rewriting bits,
demonstrating how the life of a Maxwellian demon, who utilizes the potential provided by fluctuations in a system, 
turns out to be even harder than one might expect, since any computation done by the demon based on observing such fluctuations has to 
be executed in time intervals that are of the same order of magnitude or faster than the typical lifetime of a fluctuation exploitable by the demon.

Interestingly, the isothermal action for Szilard's one atom gas bit, using a hydrogen atom, with 
$L_1\sim120\cdot10^{-12}{\rm m}$, being twice the hydrogen atom diameter, $L_2=L_1/2$, $m\sim1.7\cdot10^{-27}{\rm Kg}$, 
the Boltzmann constant $k\sim 1.4\cdot 10^{-23}{\rm JK^{-1}}$ and $T=300{\rm K}$, has the value: 
\begin{equation}
h^{\rm H}_{\rm isothermal}\sim 8.9\cdot10^{-34}{\rm Js}\,,
\label{isoactcyc2}
\end{equation}
\\
\noindent
which is of the same magnitude as Planck's constant, $h\sim 6.6\cdot10^{-34}Js$.
For Szilard's device using an iron atom with an empirical diameter of $\sim 250\cdot10^{-12}{\rm m}$ and a mass of approximately $56$ 
times the Hydrogen mass one estimates
\begin{equation}
h^{\rm Fe}_{\rm isothermal}\sim 2.8\cdot10^{-32}{\rm Js}\,.
\label{isoactcyc3}
\end{equation}
This means that theoretically a computing device with bytes of 32 Szilard ``iron-bits'' allows us to 
manipulate an approximate number of 1.1 Mbytes, rewriting each bit at a frequency of 1 Tflop ($10^{12}{\rm Hz}$) 
with a lower bound of $1{\rm W}$ irreversible power consumption at room temperature ($27^\circ{\rm C}$).
This power of $1{\rm W}$ has to be spent on fast bit switching. In comparison the power that additionally has 
to be spent on Landauer's bond, in this example only amounts to approximately $0.1{\rm W}$.
 
In contrast to microscopic machines, such as Szilard's ``one particle gases'', macroscopic devices,
bicycle pumps for instance, also require macroscopically observable amounts of irreversible work for being 
driven in finite time.
A bicicle pump with $L_1=2L_2\sim 20{\rm\ cm}$ at a temperature of $T\sim 300 {\rm\ K}$ and a cross section 
of $4{\rm\ cm}^2$ pumping Nitrogen gas has a lower bound for irreversible thermodynamic actions per particle 
in the range of $5\cdot 10^{-24}Js$ which is of an order $10^{10}$ larger than for microscopic thermodynamic cycles. 
The total thermodynamic action for the bicycle pump is then by a factor $N\sim 10^{21}$ larger, 
i.e. $Nh^{\rm bicycle-pump}_{\rm isothermal}\sim 5\cdot10^{-3} Js$.
This means that in the isothermal bicycle pump example we can go through a compression/decompression cycle 
approximately 14 times per second with a lower bound of $1W$ irreversible power-consumption. 
We point out that this irreversible work has nothing to do with friction but only with processing at finite speed.

\section{Conclusions}

We have demonstrated, using a simple gas-kinetic framework, that a cyclic thermodynamic process can be characterized by an  
action constant $h_{\rm cycle}(\theta)>0$, where $\theta$ are some parameters that characterize the process such as
particle masses $m$, dimensions of the cylinder $L$, and characteristic reference temperatures $T$. 
At the same time $h_{\rm cycle}$ does not depend on the number of particles $N$ participating in the process, 
or the cycle period $\tau_{\rm cycle}$. The thermodynamic cycles can only be driven reversibly 
in the limit of infinitely long cycle periods. 
For finite periods the amount of irreversible work per particle can not become 
smaller than $h_{\rm cycle}\tau_{\rm cycle}$. 

Moreover, we can compute the finite time Carnot cycle efficiency. In the static limit $\tau_{\rm cycle}\to\infty$
the efficiency of running a process reversibly gets recovered, while for the idle process there exists a maximum speed, 
i.e. a minimum cycle period, where the efficiency of the Carnot-process vanishes ($\eta=0$).

If we apply our result to models of computing devices, such as Szilard's one atom gas bits, then 
it follows that the minimal amount of irreversible work cannot be discussed independently from 
how frequently a bit gets reset. The theoretical lower bounds for irreversible energy consumption for embodied bits, 
such as Szilard's one particle gases, yield values of $h_{\rm cycle}$ that for microscopic systems can be of the same order of magnitude, 
or even smaller than Planck's constant.  However, for macroscopic devices $h_{\rm cycle}$ and the associated irreversible energy consumption 
of a macroscopic machine also take macroscopically observable values. 

We want to point out that the thermodynamic action principle we observe here has been derived solely on the basis of
classical mechanics. Moreover, the amount of irreversible work for thermodynamic cycles 
considered in this work has nothing to do with friction (e.g. the friction between piston and cylinder)
but with driving thermodynamic cycles with finite (non vanishing) velocities. 

\vfill

\end{document}